# Manuscript

## Title

- Explaining human cooperation through a dual mechanism of individual and social learning

## Authors

Zhihao Hou[1, †], Zhikun She[1, †], Quanyi Liang[1], Qi Su[2,3,4, *] and Daqing Li[5, *]

## Affiliations

[1] School of Mathematical Sciences, Beihang University, Beijing 100191

[2] Department of Automation, Shanghai Jiao Tong University, Shanghai 200240, China.

[3] Key Laboratory of System Control and Information Processing, Ministry of Education of China, Shanghai 200240, China.

[4] Shanghai Engineering Research Center of Intelligent Control and Management, Shanghai 200240, China.

[5] School of Reliability and Systems Engineering, Beihang University, Beijing 100083, China

[†] These authors contributed equally to this work.

[*] Corresponding Author: Qi Su (Email: qisu@sjtu.edu.cn), Daqing Li (Email: daqingl@buaa.edu.cn)

## Author contributions

Conceived and designed the experiments: Z.H., Z.S., Q.L., Q.S., and D.L.

Performed the experiments: Z.H., Z.S., and D.L.

Analyzed the data: Z.H. and D.L.

Contributed materials/analysis tools: Z.H., Z.S., and Q.L.

Wrote the paper: Z.H., Z.S., Q.S., and D.L.

## Abstract

Cooperation on social networks is crucial for understanding human survival and development. Although network structure has been found to significantly influence cooperation, human experiments have observed different cooperation phenomena under similar conditions. While evidence suggests that these differences arise from human exploration, our understanding of its impact mechanisms and characteristics remains limited. Here, we seek to formalize human exploration as an individual learning process involving trial and reflection, and integrate social learning to examine how their interdependence shapes cooperation. We find that individual learning can alter neighbor imitation tendencies, and the resulting shifts in the local cooperative environment feed back into the experiential cognition that guides individual learning. This coupled dynamic makes the ability of social networks to promote cooperation largely dependent on whether individuals focus on long-term payoffs, and exhibits a series of characteristics that can explain previously unexplained and seemingly contradictory cooperation phenomena. Surprisingly, individual learning can promote cooperation more than social learning when its probability is negatively correlated with payoffs, a mechanism rooted in the psychological tendency to avoid trial-and-error when individuals are satisfied with their current payoffs. These results explain the contradictory cooperation phenomenon by accounting for decision preferences and cognitive processes underlying exploration, bridging the gap between theoretical research and reality.

## MAIN TEXT

## Introduction

Cooperation, as a prosocial behavior widely present at all levels of human interactions, plays an indispensable role in human survival and development[1]. Identifying mechanisms that support the evolution of cooperation in highly competitive environments has long been a focus of sociology and evolutionary biology[2,3]. One major mechanism is network reciprocity, which means that social networks can help cooperators interact more frequently with other cooperators and benefit from these interactions[4–7]. In contrast to the flourishing theoretical research (ranging from regular[5] to heterogeneous[8], asymmetric[9], multilayer[10], and dynamic networks[11]), experimental evidence for network reciprocity remains mixed[12–14]. While some studies show that social networks under specific conditions can promote cooperation[15], other experiments under similar conditions report the opposite findings[16–18]. This contradiction has been attributed to human exploration, suggesting that people do not make decisions solely through social learning, as most theoretical models assume[2,19], but also engage in trial-and-error behaviors that can undermine network reciprocity[15,20,21].

One critical but still unexplained aspect is how human exploration interact with social learning in cooperation. However, previous studies typically modeled human exploration as random choices between cooperation and defection with equal probability[21–23]. A potential solution to bridge the above gap is to view human exploration as an individual learning process, a fundamental method by which humans solve problems[24]. Individual and social learning characterize the predominant ways humans make decisions in unknown environments[25–27]; that is, through trial-and-error and the use of social information. Although both have received widespread attention in psychology and decision theory[28–30], individual learning has been largely absent from formal models of the evolution of cooperation. More importantly, individual learning helps explain the motivations and cognitive processes underlying human exploration, making it a suitable framework for studying social cooperation. To test this hypothesis, we propose a coupled decision model that includes: (i) individual learning[31], where players learn the benefits of different strategies through trial-and-error, and then choose an advantageous strategy based on experiential cognition; and (ii) social learning[32], where players update their strategies by imitating successful neighbors. We find that individual learning can alter neighbor imitation tendencies, and the resulting shifts in the local cooperative environment feed back into the experiential cognition that guides individual learning. This coupled dynamic makes focusing on long-term payoffs beneficial for cooperation by driving individual learning toward cooperation. Focusing on short-term payoffs drives individual learning to favor defection, ultimately triggering the collapse of cooperation. By analyzing human behavioral patterns revealed in a series of key experiments, we show that the coupled dynamic described above may represent the underlying mechanism through which social networks shape cooperation. Surprisingly, individual learning as an exploratory behavior does not necessarily suppress cooperation relative to social learning, contrary to the intuition in previous studies[20,21]. When its probability is negatively correlated with payoffs, individual learning can even lower the critical threshold required for cooperation to prevail over defection. Our work explains the paradoxical cooperation phenomena observed in human experiments by incorporating decision preferences and the cognitive processes underlying exploration, thereby deepening our understanding of the interaction between individual and social learning in shaping cooperation.

## Results

**Individual-social coupled decision model.** We consider an evolving population of $N$ players interacting as described by a regular graph (Fig. 1). Players choose to cooperate or defect in interactions with their neighbors and receive payoffs at each time step depending on their strategies. If a player has $n_C(t)$ cooperators among its $k$ neighbors, its payoff for cooperating is $bn_C(t) - ck$, while the payoff for defecting is $bn_C(t)$. After players receive payoffs, a random player is selected to update its strategy. Here, we consider the two fundamental ways in which humans make decisions in unknown environments: trial-and-error (individual learning) and imitation of neighbors (social learning). The probability of individual learning, denoted as $P^{\mathrm{IL}} \in [0,1]$, captures how players choose between these two ways. Social learning, as the cornerstone of cultural evolution and social transmission[30,32], has been widely studied in evolutionary game theory[19]. Individual learning formalizes human exploratory behaviors centered on trial-and-error, such as heuristic strategies and various reinforcement learning algorithms[33]. Given the social context we focus on (binary state space, dynamic and repeated interactions), simple trial-and-reflection may be more applicable than other complex learning mechanisms[34,35]. The detailed decision process is as follows.

In social learning, the payoff $\pi(t)$ is translated into a social value, $f(t) = 1 - \omega + \omega\pi(t)$, where $\omega \geq 0$ measures the importance of payoffs. This transformation is well known from a genetic perspective[36], where $f$ represents fitness and $\omega$ is interpreted as the intensity of selection. We primarily focus on the weak selection[19,37], i.e., $0 < \omega \ll 1$. If the selected player chooses social learning with probability $1 - P^{IL}$, its probability of choosing cooperation is determined by the social value $f(t)$ of all cooperator neighbors $F_C(t)$, and the total social values of defector neighbors $F_D(t)$, given by $F_C(t)/(F_C(t) + F_D(t))$.

If the selected player chooses individual learning with probability $P^{IL}$, it tries to reverse its strategy and then engages in a trial-and-error process for $\mu$ time steps (during which its neighbors continue to update their strategies). The payoff experiences obtained during this period, $(\pi(t+1), \dots, \pi(t+\mu))$, are weighted by a set of coefficients $(\lambda_1, \lambda_2, \dots, \lambda_\mu)$ to capture arbitrary mappings between human cognition and experience:

$$C(t,\mu) = \lambda_1\pi(t+1) + \cdots + \lambda_\mu\pi(t+\mu). \quad (1)$$

For example, equal weights $(1/\mu, \dots, 1/\mu)$ imply uniform attention to all payoffs in $C(t,\mu)$, while $(0, \dots, 0,1)$ reflects a focus on the final outcome after $\mu$ steps. After trial-and-error, if $C(t,\mu)$ exceeds the original strategy's payoff $\pi(t)$, the player sticks with the new strategy. Otherwise, the player returns to the original strategy. Individual learning can be interrupted, which may correspond to impulsive re-decision or a loss of patience[38,39]. In an extreme case, a player undergoing a trial-and-error from C to D may be interrupted at time step $t$ and subsequently choose individual learning again with probability $P^{IL}$. In this case, it starts a new trial-and-error from D to C and compares $C(t,\mu)$ with $\pi(t)$ after $\mu$ steps.

**Network reciprocity from the individual-social coupled decision perspective.** The updating rule with exploratory behavior defines a non-absorbing Markov chain, which requires analyzing the steady-state average abundance of cooperators, $\langle p_C \rangle$. Cooperation prevails over defection if $\langle p_C \rangle > 0.5$, and we analytically derive a critical condition (see Methods) in close agreement with numerical simulations, which takes the following form:

$$\langle p_C \rangle > 0.5 \Leftrightarrow b > \frac{ck}{q(k-1)} + \delta \quad (2)$$

Although our critical condition applies to arbitrary values of $\mu$, $P^{IL}$, and $(\lambda_1, \ldots, \lambda_\mu)$, here we consider two representative cases to facilitate an efficient discussion of the effect of $(\lambda_1, \ldots, \lambda_\mu)$: one in which experiential cognition depends mainly on the terminal payoff (TP), and another in which all payoffs are weighted equally (EP). Cases where cognition relies on immediate payoffs can be captured by setting a small value of $\mu$. Figure 2a illustrates the phase space under TP, showing that cooperation more easily prevails over defection as players shift their focus from immediate payoffs to long-term outcomes (i.e., as $\mu$ increases). Figure 2b shows that the phase space under EP exhibits a pattern similar to that under TP. When $\mu$ is small, cooperation nearly vanishes. As $\mu$ increases, the level of cooperation $\langle p_C \rangle$ gradually rises. Furthermore, we verify the robustness of this phase space pattern by setting $\lambda_{i+1}/\lambda_i$ to different constant values. Figure 2c shows that while increasing $\lambda_{i+1}/\lambda_i$ can promote cooperation, the cooperation level primarily depends on $\mu$. These results suggest that the dependence of cooperation on $\mu$ does not rely on the specific form of $(\lambda_1, \ldots, \lambda_\mu)$.

To systematically explore the effect of $\mu$ and compare our model with the ideal case without exploration behavior (i.e., the well-known critical condition $b > ck$[5]), we plot the difference between the critical benefit required for cooperation to prevail over defection and the cost, $b^* - ck$, as a function of $\mu$ in Figure 2d. As a general trend, increasing $\mu$ leads to a decrease in $b^*$, suggesting that incorporating long-term payoffs into evaluation promotes cooperation. In particular, when players place little emphasis on immediate payoffs (e.g., $\lambda_{i+1}/\lambda_i > 1$, or $(\lambda_1, \ldots \lambda_{\mu-1}, \lambda_\mu) = (0, \ldots 0,1)$), the critical benefit $b^*$ can approach the ideal case. Figure 2e shows the effect of the individual learning probability $P^{IL}$ on cooperation. As $P^{IL}$ increases, cooperation becomes more difficult to prevail over defection, but the level of cooperation $\langle p_C \rangle$ still depends primarily on $\mu$, even when $P^{IL} \sim O(\omega)$ (Fig. S1a). These results indicate that the phase space structure is not an artifact of our model setup nor a consequence of specific parameter choices.

To understand the phase space structure, we examine the outcome of individual learning, which is the only decision pathway directly influenced by $\mu$. As shown in Fig. 3a, individual learning is asymmetric (i.e., $P^{IL}_{D\to C} \neq P^{IL}_{C\to D}$, where $P^{IL}_{D\to C}$ denotes the probability that a defector switches to cooperation through individual learning, and $P^{IL}_{C\to D}$ denotes the probability that a cooperator switches to defection) and gradually favors cooperation as $\mu$ increases. This is because strategy changes induced by trial-and-error can trigger imitation by neighbors, with the likelihood increasing with the number of neighbor updates (Fig. S1b). As $\mu$ increases, trials from defection to cooperation tend to result in more cooperator neighbors (blue), while trials from cooperation to defection lead to fewer cooperator neighbors (red), making individual learners more likely to favor cooperation. These individual learning outcomes modify the probabilities of both cooperators and defectors interacting with cooperators (Fig. 3b), thereby influencing the imitative tendency of social learning (Fig. 3c). Linking these results together reveals the coupling mechanism between individual and social learning: individual learning can alter neighbor imitation tendencies, and the resulting shifts in the local cooperative environment feed back into the experiential cognition that guides individual learning.

**Experiments related to the individual–social coupled decision model.** We now turn to experimental results to assess the explanatory power of our model in capturing human behavior in games. Given the differences in update frequencies between theoretical models and human experiments (the former typically adopts asynchronous updating to better approximate realistic decision-making[19,40], while the latter is usually constrained to synchronous updating due to laboratory conditions, implying that $\mu$ may require rescaling by $1/N$ or $k/N$, depending on the transformation logic), as well as the heterogeneity of subjects and experimental designs across studies, directly reproducing specific experimental results or comparing data may not always be meaningful. Therefore, instead of focusing on specific quantitative metrics, we highlight the fundamental characteristics shared by our model and human decision-making, which are reflected in the following three aspects: (1) exploration (i.e., individual learning) is asymmetric and can significantly influence cooperation; (2) payoff structures and social environments can influence the orientation of human exploration; (3) human behavior can be driven either by profit-seeking motives (individual learning) or by herd mentality (social learning).

Let us begin with features of social cooperation observed in human experiments, exemplified by two representative studies reporting contrasting results regarding network reciprocity. The negative result comes from Traulsen et al.[20], who report that the mutation probability from cooperation to defection, $P_{C\to D}^{MU}$, is higher than that from defection to cooperation, $P_{D\to C}^{MU}$, with the proportion of cooperators, $p_C$, steadily declining to below 0.3. Here, mutation refers to a specific form of human exploration (a player switching to a strategy different from all its neighbors, which is the only identifiable exploratory behavior under laboratory conditions). The positive result comes from Rand et al.[15], who find that $P_{D\to C}^{MU} > P_{C\to D}^{MU}$, with $p_C$ stabilizing above 0.5. Although these two experiments produce different outcomes under similar conditions (both satisfying $b/c > k$), the rich phase space of cooperation and the asymmetric exploration they jointly reveal cannot be explained by previous theoretical models (Fig. 4a). In the classical pure social learning model without exploration, the proportion of cooperators $p_C$ tends to converge to 1 (black solid line) or 0 (black dashed line). In the random exploration model, players choose between cooperation and defection with equal probability, causing $p_C$ to fluctuate around 0.5. However, such fundamental features of cooperation naturally emerge in our model (Fig. 4b). When $\mu$ is small, individual learners struggle to recognize the imitation feedback through trial and error, leading to $P_{D\to C}^{MU} < P_{C\to D}^{MU}$ and a steady decline in $p_C$ to below 0.3. In contrast, a large $\mu$ results in $P_{D\to C}^{MU} > P_{C\to D}^{MU}$ and stabilizes $p_C$ above 0.5. These results suggest that the coupling between individual and social learning can explain the long-unexplained asymmetric exploration and the divergent levels of cooperation observed under similar payoff structures.

Rand et al.[15] find that $P_{C\to D}^{MU}$ is higher when $b/c \leq k$ compared to $b/c > k$, while the opposite pattern holds for $P_{D\to C}^{MU}$. This finding suggests that human exploration is modulated by the payoff structure, i.e., exploration is more likely a cognitive process with some degree of rationality, rather than a random behavior akin to mutation or noise. Figure 4c shows that our model reproduces this pattern, which can be explained by the coupled dynamic between individual and social learning: the experiential cognition that guides individual learning toward cooperation requires imitation feedback shaped by social learning, which promotes cooperation only when $b/c > k$[5]. So-called "moody conditional cooperators" may also emerge from this coupling mechanism (Fig. 4d). Related experiments find that the probability of a cooperator cooperating in the next round increases with the proportion of cooperators among their neighbors $n_C/k$, whereas the correlation between a defector's cooperation probability and $n_C/k$ is relatively weak[17,41,42]. Previous studies attribute this phenomenon to the idea that a player's decision depends on both their previous action and the composition of their neighborhood[41]. However, this explanation does not clarify why the correlation is weaker for defectors than for cooperators. From the perspective of coupled decision-making, the imitation mechanism in social learning naturally creates a positive correlation between cooperation probability and $n_C/k$, while individual learning amplifies this effect for cooperators and attenuates it for defectors (Fig. S1c). This occurs because a higher $n_C/k$ in the reflection phase of individual learning often implies that a defector's strategy has failed to induce defection among its neighbors, leading the defector to persist in defection; or that a cooperator's strategy has successfully encouraged cooperation among neighbors, prompting continued cooperation. These findings indicate that our model can explain the puzzling cognitive characteristics of human exploration by framing it as a trial-and-error attempt to assess whether the payoff structure and neighborhood feedback (shaped by social learning) favor cooperation.

Kirchkamp and Nagel's research proposes that players do imitate others sometimes, but they seem to learn primarily from experience[16]. Specifically, they examine how the strategy in step $t$, $s_t$, own payoff information, $\Delta^{own}$, and neighbor payoff information, $\Delta^{other}$, influence the probability of cooperation in step $t+1$, $P_C(s_{t+1})$, and find that the coefficient for own payoff information, $\beta^{own}$, is larger than that for neighbor payoff information, $\beta^{other}$. Intuitively, this result seems to challenge the rationale for relying on social learning in related studies (supported by other experiments[18,43]). From the coupled decision perspective (Fig. 4e), the observed difference in correlation primarily stems from the distinct evaluation processes underlying the two learning pathways. In individual learning, one's own payoff serves as the direct evaluative criterion, resulting in a stronger correlation with strategy updates, i.e., a profit-seeking motive. In contrast, social learning relies on the majority strategy within the neighborhood, which weakens the influence of neighbor payoff information, i.e., a manifestation of herd mentality. This finding illustrates that the dual evaluation processes in our coupled decision model may reconcile seemingly conflicting observations about individuals' reliance on different types of game information.

**Adaptive individual learning probability can promote cooperation.** As shown in Figure 5a, although our coupled model (red arrow) exhibits a lower critical threshold than the random exploration model (yellow arrow), it merely approaches the ideal condition achieved by pure social learning, rather than surpassing it. This suggests that the condition for cooperation to prevail over defection is at least $b > ck$, which appears inconsistent with the widespread emergence of cooperation observed in highly connected real-world networks[44]. Here, we find that if players' willingness to risk trial and error is inversely related to their payoff (i.e., the probability of individual learning, $P^{IL}$, decreases as payoff increases), then individual learning can significantly lower the critical threshold required for cooperation to prevail over defection.

Noting that only cooperator neighbors contribute to a player's payoff, an adaptive $P^{IL}$ means that individual learning primarily occurs in social environments with few cooperators. Here, we set $P^{IL} = P_0^{IL} * (k - n_C)/k$, and expect this asymmetry to promote cooperation by disrupting the defector cluster more than the cooperator cluster. Figure 5b shows that the effectiveness of adaptive individual learning in promoting cooperation strongly depends on the value of $\mu$. When $\mu$ is small, the adaptive $P^{IL}$ can mitigate the negative effects of individual learning to some extent, but the critical benefit $b^*$ still exceeds $ck$. As $\mu$ increases, $b^*$ gradually decreases, and when $\mu = 500$, $b^*/ck$ falls below 0.5. This pattern can be intuitively understood by interpreting $\mu$ as the duration of asymmetric perturbation introduced by individual learning. Figure 5c further confirms the robustness of cooperation promotion through adaptive $P^{IL}$, where $P_0^{IL}$ has little impact on $b^*$. These results represent an idealized scenario, as our setting assumes that players do not engage in individual learning when they are surrounded by cooperators. However, Fig. S1d demonstrates that even when this restriction is partially relaxed, adaptive individual learning probabilities can still promote cooperation.

## Discussion

Previous studies show that strategy update rules significantly influence the evolutionary outcomes of structured populations[45]. Therefore, whether theoretical models can capture the key features of human decision-making has always been a focal debate in evolutionary game theory[15,20]. Most existing studies are based on the assumption of social learning[5,7,19]. However, experimental research shows that human decision-making not only depends on social learning but also exhibits exploratory behavior[20]. To understand this phenomenon, researchers have studied the impact of human exploration from the perspective of noise or mutation[21–23,46], but this perspective seems insufficient to explain the cognitive features revealed by experimental studies, such as asymmetric exploration and environmental influences. To bridge this gap, we seek to formalize human exploration as an individual learning process involving trial and reflection, and integrate social learning to examine how their interdependence shapes cooperation. We find that individual learning can alter neighbor imitation tendencies, and the resulting shifts in the local cooperative environment feed back into the experiential cognition that guides individual learning. Consequently, prioritizing long-term outcomes makes individual learners more likely to recognize this feedback and thus choose cooperation. Focusing on short-term payoffs drives individual learning to favor defection, ultimately triggering the collapse of cooperation. The key insight of these findings lies in offering a novel explanation for why individuals tend to value long-term payoffs. Social interactions can induce a correlation between cooperation and long-term orientation without relying on direct reciprocity[47,48]. To further validate our findings, we compare the characteristics of our model with human behavior observed from experimental studies, and show that the coupled dynamic between individual and social learning can capture (at least to some extent) the real mechanisms through which social networks influence cooperation. Finally, we find that individual learning can promote cooperation more than social learning if players' willingness to risk trial and error is inversely related to their payoffs. In fact, this adaptive individual learning probability is not groundless. Extensive experimental evidence suggests that when people are satisfied with the status quo, they typically do not exert additional effort to find optimal solutions, a phenomenon known as the satisficing principle[49–51].

Our work provides a more comprehensive framework for studying human behavior in game interactions, as reflected in three main contributions: (1) we reveal the coupling mechanism between individual and social learning, which not only links the cooperation differences observed in experimental studies to the previously unexplained features of human exploration, but also offers a novel explanation for the correlation between cooperation and long-term orientation without relying on direct reciprocity; (2) we show that individual learning can promote cooperation more than social learning when its probability is negatively correlated with payoffs, challenging the prevailing intuition that human exploration merely hinders cooperation[20,21]; (3) we provide a systematic theoretical framework for strategy updating mechanisms based on temporal sequences of payoffs, which is the first study to incorporate a multi-step decision process into the analysis of evolutionary dynamics. One limitation of our study is that the analysis partly relies on exogenous parameters (e.g., a fixed $\mu$). Although incorporating dynamic feedback between decision preferences and social interactions may further complicate and enrich the problem[52–55], addressing this issue remains a long-term endeavor that requires interdisciplinary collaboration. As the first step toward incorporating individual learning into the evolution of cooperation on social networks, our primary focus here is whether the coupled model can capture key features of human decision-making.

As the predominant ways in which individuals understand their environment[56], individual and social learning are prevalent across various aspects of human and animal life. For example, the learning process of human infants[57], the foraging process of animals[58,59], and other social interaction scenarios similar to the focus of this article[60]. Human infants acquire language skills through trial-and-error correction and observational imitation. Animals such as chimpanzees and wasps improve their foraging efficiency by observing the foraging behavior of their conspecifics while attempting to find optimal food sources. Moreover, a series of experimental studies show that individual and social learning play a dominant role in human cognitive formation and behavioral decisions, including everyday activities such as shopping[61], driving[62], and playing games[63]. "I help you in the hope that you will become like me, thereby improving my social environment." We believe that studying the evolution of cooperation from the individual-social coupled decision perspective offers a promising pathway to bridge the gap between theoretical models and real-world decision making.

## Materials and Methods

**Notation.** We consider a structured population of $N$ individuals consisting of cooperators (C) and defectors (D). The interaction and replacement relations between individuals are depicted by an undirected regular network $G$. On the network, each individual $i$ has $d$ neighbors. Let $Q$ be the normalized adjacency matrix of network $G$. If nodes $i$ and $j$ are connected, the weight between them is $q_{ij} = q_{ji} = 1/d$; otherwise, $q_{ij} = q_{ji} = 0$. The total weight of individual $i$ is $q_i = \sum_j q_{ij} = 1$. The state of the system can be described by a (column) binary vector $\boldsymbol{s} = (s_1, s_2, \ldots, s_N)$, $s_i \in \{0,1\}$, where 1 represents a cooperator and 0 represents a defector. Let $p_C$ and $p_D$ denote the frequencies of C and D in the population, $p_C = \frac{1}{N}\sum_i s_i$ and $p_D = 1 - p_C$. Let $p_{CC}$, $p_{CD}$, $p_{DC}$ and $p_{DD}$ denote the frequencies of CC、CD、DC and DD pairs, respectively. Let $q_{X|Y}$ denote the conditional probability to find an X-player given that the adjacent node is occupied by a Y-player. Here, both X and Y stand for C or D. The following variable relations show that the whole system can be described by $p_C$ and $q_{C|C}$.

$$
\begin{aligned}
p_C + p_D &= 1, \\
q_{C|X} + q_{D|X} &= 1, \\
p_{XY} &= q_{X|Y} * p_Y, \\
p_{CD} &= p_{DC}.
\end{aligned}
\tag{3}
$$

**Condition for cooperation to prevail over defection.** To derive the critical condition, we need to estimate the dynamical equation of $p_C$. Let $P^{SL}_{X\to Y}$ denote the probability that the selected player transitions from X to Y through social learning, and $P^{IL}_{X\to Y}$ denote the probability that the selected player transitions from X to Y through individual learning, satisfying $P^{SL}_{X\to X} + P^{SL}_{X\to Y} = 1$ and $P^{IL}_{X\to X} + P^{IL}_{X\to Y} = 1$. Then, the time derivative of $p_C$ is given by

$$
\begin{aligned}
\dot{p_C} = &\frac{(1 - P^{IL})}{N}\left(p_D P^{SL}_{D\to C} - p_C P^{SL}_{C\to D}\right) + \frac{P^{IL}}{N}(1 - 2p_C) \\
&+ \frac{P^{RE}}{N}\left(p_D P^{IL}_{D\to C} - p_C P^{IL}_{C\to D}\right) \triangleq F(p_C, \omega) + O(\omega^2).
\end{aligned}
\tag{4}
$$

Here, $P^{RE} = P^{IL}\left(1-\frac{1}{N}\right)^{\mu}$ denotes the probability that individual learning is not interrupted, and $F(p_C,\omega)$ gives the dynamical equation after neglecting terms of order $O(\omega^2)$. In Eq. (4), the first term on the right-hand side of the first equality represents the impact of social learning, the second term captures the effect of the trial-and-error process in individual learning, and the third term reflects the influence of the reflection process in individual learning.

**Social learning — single-step decision process.** In social learning, players rely only on the strategies and payoffs of their neighbors in the current step, meaning that social learning is a single-step decision process. To calculate the effect of social learning, we need to estimate the neighborhood distribution and the payoffs of neighbors, thereby determining $P^{SL}_{D\to C}$ and $P^{SL}_{C\to D}$ (for details, see Chapter 3.1 of the Supporting Information). Under weak selection, we obtain

$$p_D P^{SL}_{D\to C} - p_C P^{SL}_{C\to D} = \frac{\omega p_{CD}}{k}(k-1)\big(b(k-1)\big(q_{C|C}-q_{C|D}\big)-ck\big)\big(q_{C|C}+q_{D|D}\big)+O(\omega^2). \tag{5}$$

**Individual learning — multi-step decision process.** After choosing individual learning, players undergo a trial-and-error phase lasting $\mu$ time steps and reflect (update their strategy again) at the end based on the payoffs obtained during this period, $\big(\pi(t+1),\pi(t+2),\dots,\pi(t+\mu)\big)$; that is, individual learning is a multi-step decision process. Note that under weak selection, we have

$$\begin{aligned} P^{RE}\big(p_D P^{IL}_{D\to C} - p_C P^{IL}_{C\to D}\big) &= P^{RE}\big(p_D P^{IL}_{D\to C|\omega=0} - p_C P^{IL}_{C\to D|\omega=0}\big) \\ &+ P^{RE}\omega\left(p_D\frac{\partial P^{IL}_{D\to C|\omega=0}}{\partial\omega} - p_C\frac{\partial P^{IL}_{C\to D|\omega=0}}{\partial\omega}\right)+O(\omega^2). \end{aligned} \tag{6}$$

To calculate $P^{IL}_{D\to C}$ and $P^{IL}_{C\to D}$, we need to estimate the state transitions in the number of cooperators within the individual learner's neighborhood during its trial-and-error period, which follows the transition matrix below

$$P_\Omega = \begin{pmatrix} p_{00} & \cdots & p_{k0} \\ \vdots & \ddots & \vdots \\ p_{0k} & \cdots & p_{kk} \end{pmatrix}. \tag{7}$$

Here, $p_{n_C(n_C+1)}$ denotes the probability that the number of cooperators in the individual learner's neighborhood transitions from $n_C$ to $n_C+1$. Let $P(t) = (p_0^t,\dots,p_k^t)$ represent the probability distribution over the number of cooperator neighbors at time $t$ (i.e., before the onset of individual learning). For the explicit expressions of $P_\Omega$ and $P(t)$, see Chapter 3.2 of the Supporting Information. The distribution at time $t+\Delta t$ is given by $P(t+\Delta t) = \big(p_0^{t+\Delta t},\dots,p_k^{t+\Delta t}\big) = P(t)P_\Omega^{\Delta t}$. According to Eq. (7), we have

$$\begin{aligned} p^{IL}_{D\to C}(\mu,\lambda_1,\dots,\lambda_\mu) &= E\left[g_*\left(\lambda_1\pi(t+1)+\cdots+\lambda_\mu\pi(t+\mu)-\pi(t)\right)\right] \\ &= \sum_{\left(n_C^t,\dots,n_C^{t+\mu}\right)\in\mathcal{N}_C} g_*\left(\sum_{\Delta t=1}^{\mu}\lambda_{\Delta t} b n_C^{t+\Delta t} - b n_C^t - ck\right) p_{n_C^t}\prod_{\Delta t=0}^{\mu-1} p_{n_C^{t+\Delta t} n_C^{t+\Delta t+1}}\,. \end{aligned} \tag{8}$$

$$p_{C\to D}^{IL}(\mu,\lambda_1,\dots,\lambda_\mu) = E\left[g_*\left(\lambda_1\pi(t+1)+\cdots+\lambda_\mu\pi(t+\mu)-\pi(t)\right)\right]$$
$$= \sum_{\left(n_C^t,\dots,n_C^{t+\mu}\right)\in\mathcal{N}_C} g_*\left(\sum_{\Delta t=1}^{\mu}\lambda_{\Delta t} b n_C^{t+\Delta t} - b n_C^t + ck)\right) p_{n_C^t}\prod_{\Delta t=0}^{\mu-1} p_{n_C^{t+\Delta t} n_C^{t+\Delta t+1}}\,. \tag{9}$$

Here, $\mathcal{N}_C$ denotes the state space of the number of cooperators within the individual learner's neighborhood during its trial-and-error period, and the size of this state space is $M = (k+1)^\mu$.

**Coupling between individual and social learning.** Eq. (4) appears to suggest that social learning and individual learning are two separate processes, but this is not the case. On the one hand, social learning influences the transition matrix $P_\Omega$ (as the state transitions of neighbors depend on both social and individual learning), which is required to calculate the effect of individual learning. On the other hand, individual learning modulates the conditional probabilities $q_{C|C}$ and $q_{D|D}$, which are necessary to calculate the effect of social learning. To clarify the latter, we use the idea of identity by descent (IBD)[64]. Two individuals are IBD if no mutation separates either of them from their common ancestor. Let $q_{ij}$ denote the stationary probability that the occupants of nodes $i$ and $j$ are IBD to each other under $\omega = 0$, which satisfies the following linear recurrence relations

$$q_{ij} = \begin{cases} 1 & i=j \\ \dfrac{1-P_1^{IL}}{2}\displaystyle\sum_{k\in G}\left(p_{ik}q_{jk}+p_{jk}q_{ik}\right) & i\neq j \end{cases}, \tag{10}$$

where $P_1^{IL} = \frac{NP^{IL}+(N-2)P^{RE}}{N+(N-2)P^{RE}}$. We define $q = \frac{1}{N}\sum_{i,j} p_{ij}q_{ij} = \sum_j p_{ij}q_{ij}$ as the stationary average IBD probability among nodes. According to Lemma 1 of the Supporting Information, we have

$$P\left(s_j = 1\middle|s_i = 1\right) = q_{ij} + \left(1-q_{ij}\right)p_C = q_{ij}(1-p_C) + p_C,$$
$$P\left(s_j = 0\middle|s_i = 0\right) = q_{ij} + \left(1-q_{ij}\right)(1-p_C) = q_{ij}p_C + 1 - p_C,$$

and

$$q_{C|C} = \sum_j p_{ij}P\left(s_j = 1\middle|s_i = 1\right) = (1-p_C)q + p_C,$$
$$q_{D|D} = \sum_j p_{ij}P_{\omega=0}\left(s_j = 0\middle|s_i = 0\right) = p_C q + 1 - p_C. \tag{11}$$

**Critical condition.** According to Chapter 3.4 of the Supporting Information, we obtain $\langle p_C\rangle > 0.5$ if and only if $F(0.5,\omega) > 0$. Using Eqs. (5) and (11), we obtain the condition for cooperation to prevail over defection:

$$\langle p_C\rangle > 0.5 \Leftrightarrow b > \frac{ck}{q(k-1)} + \delta. \tag{12}$$

Here, $q$ is the stationary average IBD probability among nodes, $\delta = 4kP^{RE}(p_D P_{D\to C}^{IL} - p_C P_{C\to D}^{IL})_{|*}/[\omega(k-1)^2(1-P^{IL})(1-q^2)]$, where the subscript "$|*$" indicates that the variable is evaluated using Eq. (6) under $p_C = 0.5$ and neglecting terms of order $O(\omega^2)$. According to Eq. (18) of the Supporting Information, $q$ approaches $1/(k-1)$ as $P^{IL} \to 0$, which allows us to recover the rule $b > ck$[5]. For a more detailed procedure and the derivation under the adaptive individual learning probability, see the Supporting Information.

**Simulations.** In this study, we primarily focus on random regular networks. The data in the figures are mainly collected from 200 independent runs on random regular networks. Each network runs for $4000 \times N$ time steps (excluding the time shown on the axes), with the first $2000 \times N$ time steps considered transient and the last $2000 \times N$ time steps considered the steady state (i.e., the data collection phase).

**Acknowledgments.** We thank Boyu Zhang and Zhigang Cao for their valuable comments on the manuscript. This work is supported by the National Key R&D Program of China with grant number 2022YFA1005103 and the National Natural Science Foundation of China with grant number 12371452, 72225012 and 72288101.

**Competing interests.** The authors declare that they have no competing interests.

**Data and materials availability.** Source data are provided with this paper.

**Code availability.** All numerical calculations and computational simulations were performed in Julia. The program code is publicly available at Code Ocean or https://github.com/ciao621/Individual-and-social-learning.

# Figures

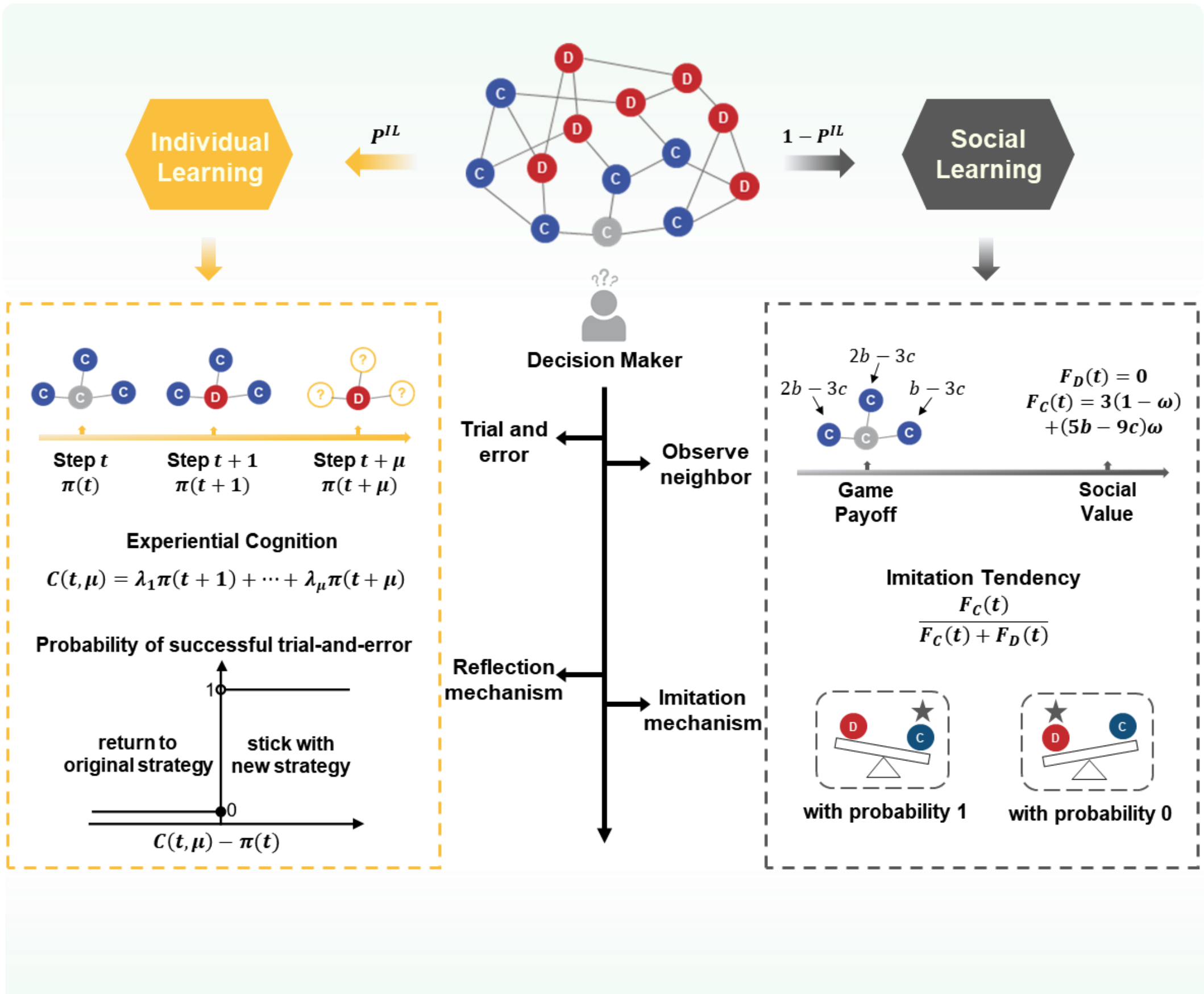

**Fig. 1 | Individual-social coupled decision model.** Red nodes represent defectors (D), blue nodes represent cooperators (C), and edges represent social interactions. At each time step, a randomly selected player (gray) updates their strategy through individual learning (yellow, with probability $P^{IL}$) or social learning (dark gray, with probability $1-P^{IL}$). Social learning is a single-step decision process, as players rely only on the strategies and payoffs of their neighbors in the current step. In individual learning, players undergo a trial-and-error phase lasting $\mu$ time steps and reflect (update their strategy again) at the end based on the payoffs obtained during this period, $\left(\pi(t+1),\pi(t+2),\ldots,\pi(t+\mu)\right)$, i.e., individual learning is a multi-step decision process.

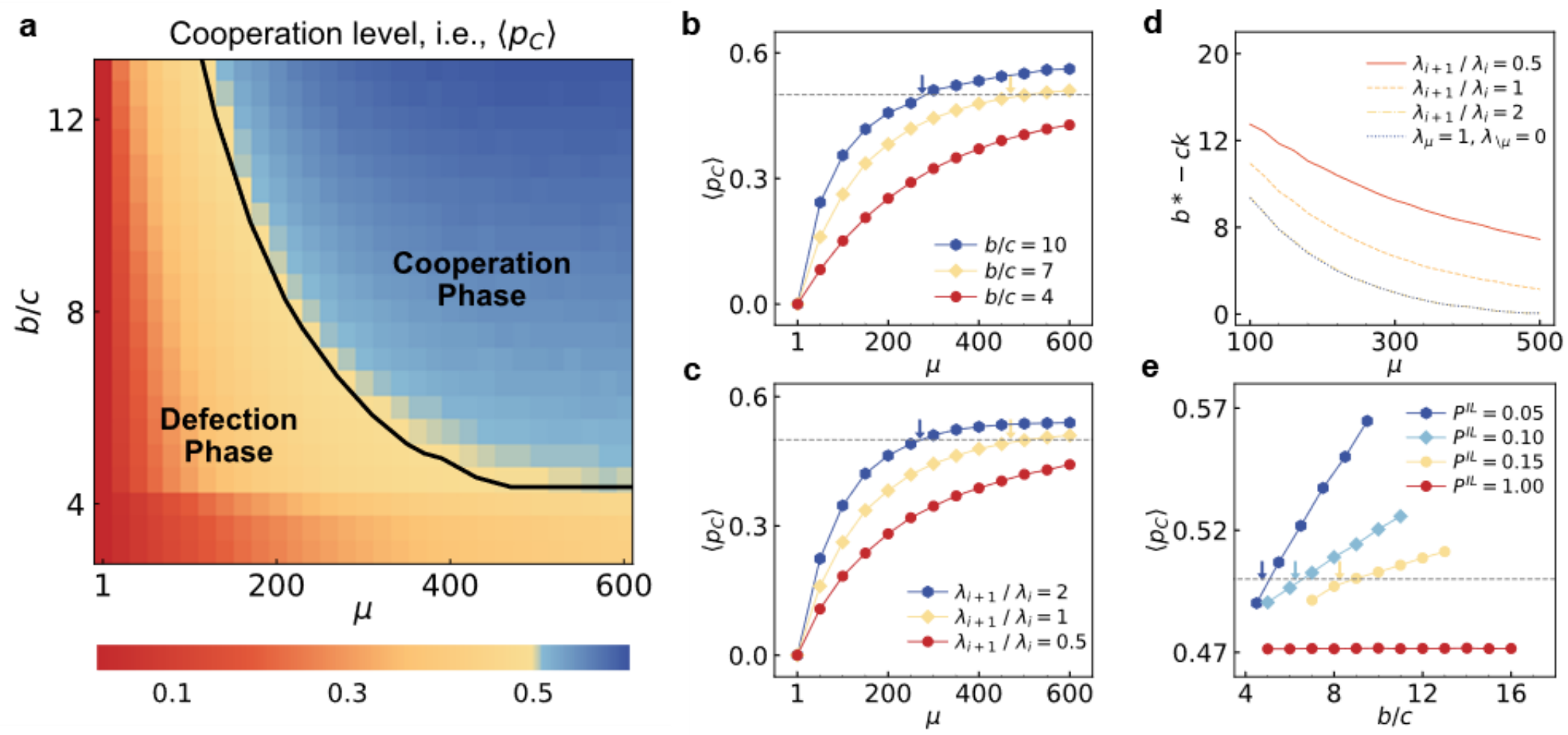

**Fig. 2 | Network reciprocity from the individual-social coupled decision perspective.** In panel a, the black line denotes the analytically derived critical condition. In panels b, c, and e, arrows indicate the analytically derived critical condition, and the grey line marks $\langle p_C \rangle = 0.5$. **a:** Phase diagram of social cooperation under $\left(\lambda_1, \dots, \lambda_{\mu-1}, \lambda_\mu\right) = (0, \dots, 0, 1)$, with colors representing the cooperation level, $\langle p_C \rangle$. **b:** Cooperation level as a function of $\mu$ under $\left(\lambda_1, \dots, \lambda_\mu\right) = (1/\mu, \dots, 1/\mu)$, with different values of $b/c$ indicated in the legend. **c:** Cooperation level as a function of $\mu$ under $b/c = 7$, with specific ratios shown in the legend. **d:** The critical condition for cooperation to prevail over defection, $(b/c)^*$, plotted as a function of $\mu$ based on our theoretical results. **e:** Cooperation level as a function of $b/c$ under $\left(\lambda_1, \dots, \lambda_{\mu-1}, \lambda_\mu\right) = (0, \dots, 0, 1)$ and $\mu = 400$, with different values of $P^{IL}$ indicated in the legend. Parameters: $P^{IL} = 0.05$ (except in panel e), $c = 1$, $k = 4$, and $\omega = 0.01$.

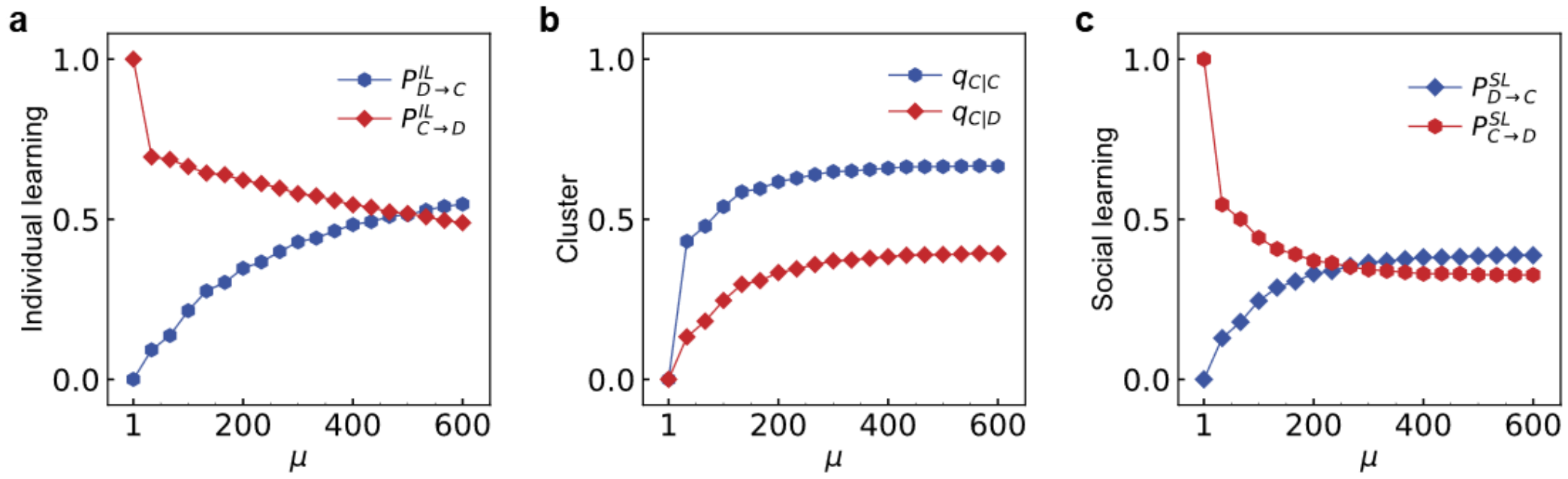

**Fig. 3 | The coupling mechanism between individual and social learning. a:** Outcome of individual learning as a function of $\mu$, where $P^{IL}_{D\to C}$ denotes the probability that a defector switches to cooperation through individual learning (i.e., from defection before trial-and-error to cooperation after reflective updating), and $P^{IL}_{C\to D}$ denotes the probability that a cooperator switches to defection through individual learning. **b:** Cluster as a function of $\mu$, where $q_{C|C}$ represents the average fraction of a cooperator's cooperator neighbors and $q_{C|D}$ represents the average fraction of a defector's cooperator neighbors. **c:** Outcome of social learning as a function of $\mu$, where $P^{SL}_{D\to C}$ denotes the probability that a defector switches to cooperation through social learning, and $P^{SL}_{C\to D}$ denotes the probability that a cooperator switches to defection through social learning. Parameters: $(\lambda_1, \dots, \lambda_{\mu-1}, \lambda_\mu) = (0, \dots, 0, 1)$, $P^{IL} = 0.05$, $b = 7$, $c = 1$, $k = 4$, and $\omega = 0.01$.

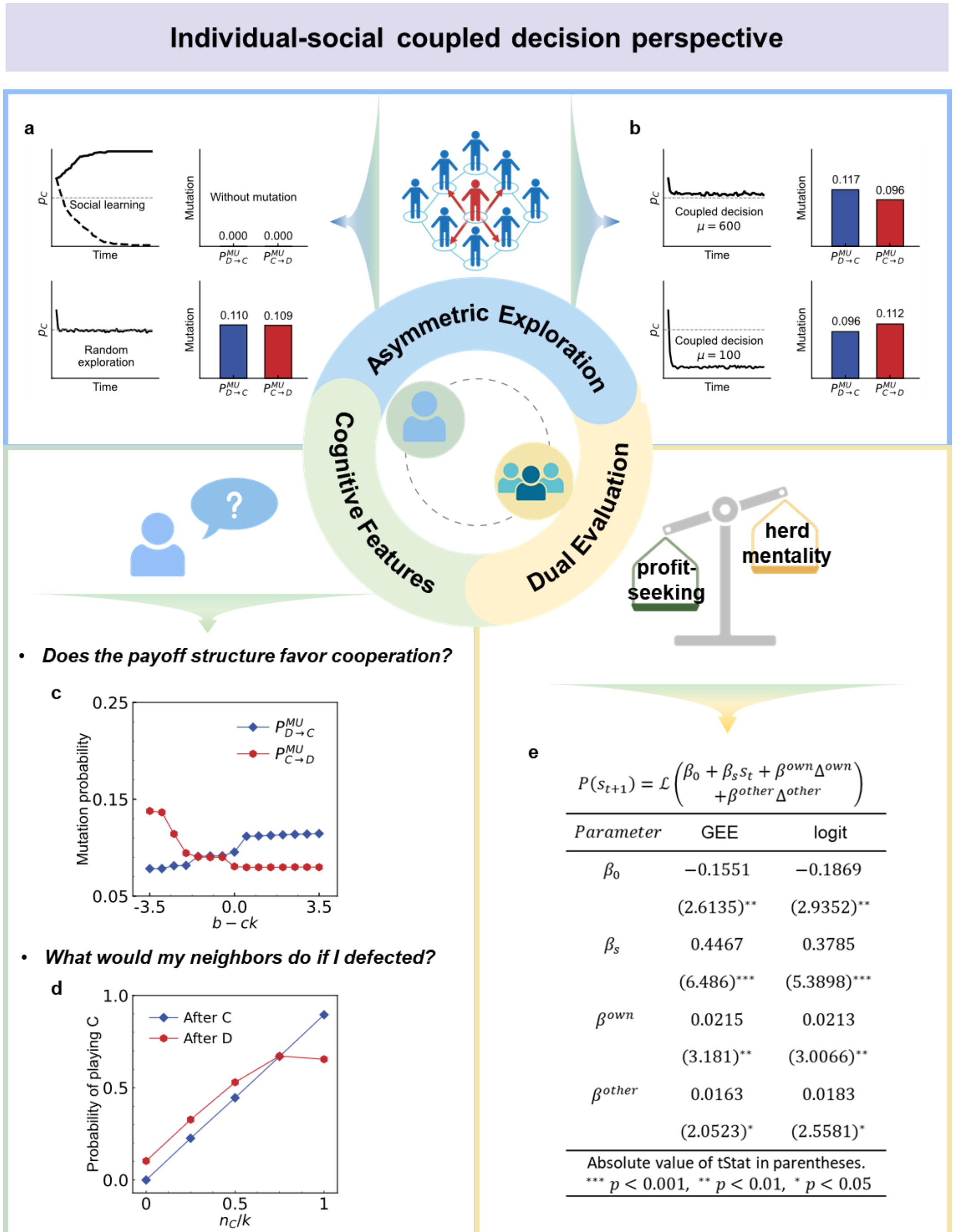

**Fig. 4 | The correspondence between the coupled decision model and human behavior. a-b:** Differences between theoretical models regarding cooperation evolution and exploration orientation. The thin gray dashed line marks $p_C = 0.5$. **c:** Mutation probability as a function of $b/c$ under $\mu = 600$. **d:** Probability that a player continues to cooperate in the next round as a function of the proportion of cooperators among its neighbors, $n_C/k$, under $\mu = 100$. **e:** Correlation between strategy updating and own versus neighbor payoffs, based on data from our model with $\mu = 100$. Parameters: $(\lambda_1, \dots, \lambda_{\mu-1}, \lambda_\mu) = (0, \dots, 0, 1)$, $P^{IL} = 0.1$, $b = 7$ (except in panel c), $c = 1$, $k = 4$, and $\omega = 0.01$.

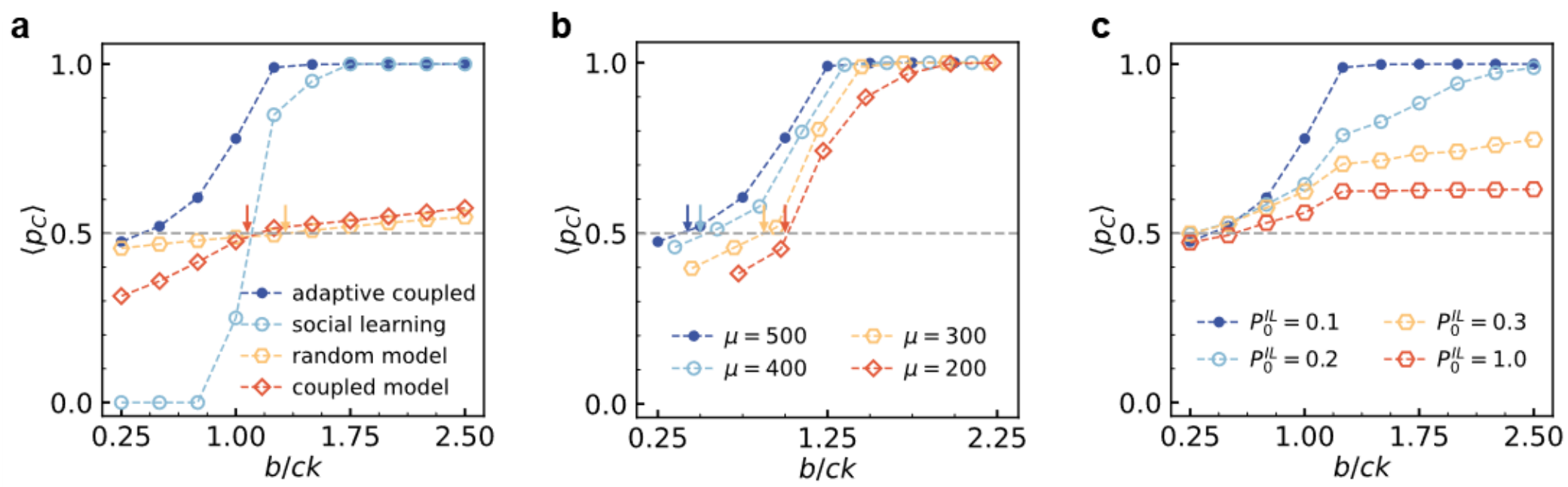

**Fig. 5 | Adaptive individual learning probability can promote cooperation.** Arrows indicate the critical condition derived analytically, and the gray line marks $\langle p_C \rangle = 0.5$. **a:** Cooperation level as a function of $b/c$, with model types indicated in the legend. Except for the pure social learning model, all models share the same mutation probability to eliminate the effect of differing exploration frequencies. **b:** Cooperation level as a function of $b/c$ under adaptive individual learning probability, with different values of $\mu$ indicated in the legend. **c:** Cooperation level as a function of $b/c$ under $\mu = 500$, with different values of $P_0^{IL}$ indicated in the legend. The theoretical results are omitted in this figure to avoid visual clutter caused by closely spaced arrows. Parameters: $(\lambda_1, \dots, \lambda_{\mu-1}, \lambda_\mu) = (0, \dots, 0, 1)$, $P_0^{IL} = 0.1$ (except in panel c), $c = 1$, $k = 4$, and $\omega = 0.01$.